\documentstyle[prl,twocolumn,aps,graphicx]{revtex}

\begin{document}

\draft


\twocolumn
[\hsize\textwidth\columnwidth\hsize\csname@twocolumnfalse\endcsname
\title{Coupled Magnetic Excitations in Single Crystal PrBa$_2$Cu$_3$O$_{6.2}$}

\author{S.J.S. Lister$^1$, A.T. Boothroyd$^1$, N. H. Andersen$^2$, B. H. Larsen$^2$, A. A. Zhokhov$^3$, A. N. Christensen$^4$, A. R. Wildes$^5$}

\address{$^1$ Department of Physics, Oxford University, Clarendon Laboratory, Parks Road, Oxford, OX1 3PU,
United Kingdom\\ $^2$ Ris{\o} National Laboratory, DK--4000
Roskilde, Denmark\\ $^3$ Russian Academy of Sciences, Institute of
Solid State Physics, Chernogolvka 14232, Russia\\ $^4$ University
of {\AA}rhus, Nordre Ringgade 1, DK--8000 {\AA}rhus C, Denmark
\\ $^5$ Insitut Laue-Langevin, B.P. 156--38042 Grenoble Cedex 9,
France}


\date{\today}
\maketitle

\begin{abstract}
The dispersion of the low-energy magnetic excitations of the Pr
sublattice in PrBa$_2$Cu$_3$O$_{6.2}$ is determined by inelastic
neutron scattering measurements on a single crystal. The
dispersion, which shows the effect of interactions with the Cu
spin-waves, is well described by a model of the coupled Cu--Pr
magnetic system. This enables values for the principal exchange
constants to be determined, which suggest that both Pr--Pr and
Cu--Pr interactions are important in producing the anomalously
high ordering temperature of the Pr sublattice. Measurements of
the Cu optic spin wave mode show that the inter-layer Cu--Cu
exchange is significantly lower than in YBa$_2$Cu$_3$O$_{6.2}$.
\end{abstract}
 ]

The anomalous properties of non-superconducting
PrBa$_2$Cu$_3$O$_{6+x}$ (PrBCO) \cite{radousky} remain an
outstanding problem in the field of cuprate superconductivity. Its
magnetic properties are strikingly different from those of other
members of the RBCO family (where R is a rare earth, Y or La).
Long range antiferromagnetic order in both the Cu and Pr
sublattices along with semiconducting resistivity persist over the
entire range of oxygen doping $0<x<1$, and the magnetic transition
in the Pr sublattice occurs at a temperature $T_{\mathrm Pr}$ an
order of magnitude higher than for other rare earths in RBCO
\cite{allenspach}. It is now generally believed that these
anomalous properties are caused by a hybridisation of the Pr $4f$
and O $2p$ orbitals. In particular, an influential model of the
electronic structure of PrBCO \cite{FR-PRL-1993} explains the
absence of superconductivity by a localisation of the holes in
hybridised Pr--O bonds. Given the likelihood that the electronic
and magnetic properties of PrBCO, including the occurrence of
superconductivity \cite{scpbco}, both depend on this
hybridisation, detailed information on the strengths and
symmetries of the magnetic interactions is desirable.

In this Letter, we describe inelastic neutron scattering
measurements of the magnetic excitations of PrBCO made for the
first time on a single crystal. Using this data and a spin wave
model of the coupled Cu--Pr system we have independently
determined the magnitudes of the Pr--Pr, Cu--Pr and Cu--Cu
interactions. We find that the Pr--Pr interaction is unexpectedly
strong and would alone lead to $T_{\mathrm Pr}$ being
substantially higher than the rare-earth ordering temperature in
other RBCO. We have also identified the effects of a
pseudo-dipolar component to the Cu--Pr coupling on the magnetic
excitations. Finally, we show that the inter-layer Cu--Cu exchange
is reduced by a factor of two in PrBCO relative to YBCO.

The experiments were performed on a crystal of
PrBa$_2$Cu$_3$O$_{6.2}$ of mass 2\,g prepared by top seeding a
flux. The crystal mosaic was $\sim1^\circ$ and $T_{\mathrm
Pr}=13$\,K. Measurements on the triple-axis spectrometers TAS\,6
at Ris{\o} and IN8 at the ILL were made at energy transfers up to
10\,meV and 70\,meV respectively. A horizontally-focussing
analyser was employed to increase the measured signal at the
expense of some resolution in the scattering vector {\bfseries Q}.
The crystal was mounted in a $^{4}$He cryostat and measurements
were made in the $(h,k,0)$ or $(h,h,l)$ scattering planes.

Previous neutron scattering measurements on polycrystalline
samples of PrBCO$_{6}$ revealed two peaks centred at 1.7 and
3.4\,meV at $T=5$\,K, which shifted to lower energies as
$T_{\mathrm Pr}$ was approached \cite{jostarndt}. These peaks
correspond to excitations of the Pr ions subject to the local
exchange and crystal fields. To investigate the ${\bf Q}$
dependence of these excitations we performed energy scans at
$T=1.8$\,K at points in reciprocal space primarily along the
principal symmetry directions of the magnetic zone. Fig. 1 shows
representative scans at ${\bf Q}=(0,0,2.2)$ and $(0.75,0.75,0)$.
Two peaks, labelled A and B, are observed at energies consistent
with the polycrystalline data. Fig. 1 also illustrates the
variation in intensity of the excitations with the direction of
{\bfseries Q}. The intensity of peak A is largest when {\bfseries
Q} is in the $ab$ plane, and is reduced for $\bf{Q} \| c$. Peak B
exhibits the opposite trend. The width of both peaks (1-2\,meV
FWHM) is larger than the energy resolution of the spectrometer
(0.4\,meV FWHM). The implied intrinsic broadening is consistent
with relaxation due to hybridisation as suggested earlier from the
spectrum of polycrystalline PrBCO \cite{jostarndt}. The use of a
single crystal, however, shows that the observed broadening is not
solely due to dispersion.

Single crystals also enable us to study the dispersion of the
excitations, which is a central objective of this work. The insets
to Fig. 1 show the low-energy data corrected for the non-magnetic
background and elastic peak measured on a similar-sized crystal of
YBCO$_{6.2}$. The line shows a function comprising two damped
harmonic oscillator response functions (including detailed balance
factor) fitted to the data. No variation in the energy of peak B
was apparent to within the experimental uncertainty. The energy
dispersion of peak A was obtained by fitting a single response
function to the data for $E\leq3$\,meV. The results are shown in
Fig. 2. A number of scans were also made at ${\bf Q}=(h, h, l)$
for $l\neq0$, but no measurable dispersion along the $c$-axis was
observed.

We also studied the excitations of the Cu sublattice. The Cu spin
waves in antiferromagnetic YBCO have been described by Tranquada
{\itshape et al.} \cite{tranq}.  Due to the bilayer structure of
YBCO the spectrum consists of two acoustic and two optic modes,
each pair being split into in- and out-of-plane modes by the
anisotropy (a few meV in YBCO). The optic mode is of interest
because its energy depends on the exchange coupling between
adjacent layers, which could be sensitive to the electronic
structure of the Pr ion. To measure the optic mode at the magnetic
zone centre a number of constant energy scans were made in the
$(h,h,0)$ direction through the point
$(\frac{1}{2},\frac{1}{2},6.75)$, close to where the optic mode
has maximum intensity. Scans at 45 and 55\,meV are shown in Fig.
3(a). All scans at energies above $\sim$50\,meV have a peak
centred at $h=\frac{1}{2}$, which we interpret as being due to the
optic mode. The amplitude of this peak exhibits a stepwise
increase with energy, as shown in Fig. 3(b). The optic mode gap,
taken from the mid-point of the step, is $53\pm2$\,meV. The
corresponding energy for YBCO$_{6.2}$ estimated in the same way
from the available data \cite{optics} is $70\pm5$\,meV.

We now consider a model to describe the dispersion of peak A. The
model is based on the single-ion states of Pr$^{3+}$ in the
tetragonal crystalline electric field (CEF). We obtained these
states by refining a model for the CEF using the observed energy
spectra of polycrystalline PrBCO$_6$ \cite{Hilscher}. The
refinement gave a doublet ground state with a first excited state
a few meV higher in energy, consistent with previous analyses
\cite{jostarndt,Hilscher}. To take into account the magnetic
ordering at $T<T_{\mathrm Pr}$ we included a molecular field. This
field lifts the degeneracy of the ground-state doublet, hence we
interpret peak A in Fig. 1 as the transition within the
exchange-field-split doublet. Peak B is then the transition from
the ground state to the crystal field singlet. We chose the
magnitude and direction of the molecular field to reproduce the
average energy of peak A and the observed direction of the moment
\cite{ATBPRL97} at about $45^\circ$ to the $c$ axis \cite{axes}.
The observed variation of the relative intensity of peaks A and B
with the direction of {\bfseries Q} is consistent with the
cross-sections calculated from the CEF wavefunctions.

The simplest model for the dispersion of peak A is then given by
the Hamiltonian:
\begin{equation}
H_{\rm Pr} = H_{\rm CEF}+\frac{1}{2}J_{\rm Pr}\sum_{\langle
ij\rangle} ({\bf J}_i-\langle{\bf J}_i\rangle)\cdot({\bf
J}_j-\langle{\bf J}_j\rangle) \label{eq:Hpp}
\end{equation}
where $J_{\rm Pr}$ represents an isotropic exchange interaction
between nearest-neighbour Pr ions (assumed to be $>0$ consistent
with the observed antiferromagnetic ordering of the Pr moments),
${\bf J}_i$ is the angular momentum of a Pr ion and $H_{\rm CEF}$
is the single-ion crystal field Hamiltonian including the
molecular-field term. The summation in (\ref{eq:Hpp}) is over all
pairs of nearest-neighbouring Pr ions. We use the pseudo-boson
approximation \cite{buyers} to calculate the dispersion from
(\ref{eq:Hpp}). This method gives a simple relation between matrix
elements of the calculated crystal field states and operators
corresponding to deviations of the angular momentum from the
mean-field values. Here we only include the transition within the
doublet ground state, as the exchange coupling does not give
appreciable mixing with the singlet level. The resulting
dispersion of the doublet transition has two branches (due to the
antiferromagnetic order) whose degeneracy is lifted by the
non-zero $x$ component of the exchange field. In practice, only
one of the two modes is actually observed at a given {\bfseries Q}
owing to the variation in intensity between magnetic zones. The
higher energy mode has a much larger cross-section near magnetic
reciprocal lattice vectors that coincide with nuclear Bragg
reflections (e.g. $(1,1,0)$) whereas the lower energy mode has
maximum intensity at the reciprocal lattice points corresponding
to magnetic-only reflections (e.g. $(\frac{1}{2},\frac{1}{2},0)$).
This intensity variation is in agreement with that observed
experimentally, providing further confirmation for the
identification of peak A with the intra-doublet transition, and
for the assumption of an exchange field at an angle to the $c$
axis.

As a first approximation, we then attempt to fit this dispersion
relation to the experimental data by varying $J_{\rm Pr}$. The
best fit is given by $J_{\rm Pr}=0.029$\,meV, and is shown by the
broken line in Fig. 2. Although the general form of the dispersion
is reasonably well described by this model involving only the Pr
sublattice, there is clear evidence of a discrepancy near the
magnetic zone centre. This deviation suggests that we need to
consider the effect of interactions between the Pr excitations and
the spin-wave excitations of the Cu sublattice. Owing to the very
high Cu spin-wave velocity, these interactions are only important
close to the magnetic zone centres at low energies. We model the
Cu--Pr coupling with the same anisotropic term that has been found
necessary to explain the observed non-collinear magnetic structure
\cite{ATBPRL97,maleev,ATB_Rcu}.
\begin{equation}
H_{\rm Cu-Pr} = \frac{1}{2}\sum_{\alpha\beta,\langle ij\rangle}
K_{\alpha\beta}(ij) S_{\alpha i} J_{\beta j} \label{eq:Hcupr}
\end{equation}
where $\alpha,\beta$ label components $\{x,y,z\}$, ${\bf K}(ij)$
is a tensor describing the pseudo-dipolar interaction (consistent
with the tetragonal symmetry of the crystal), and {\bfseries S}
denotes the spin of a Cu ion. A similar term was also used to
describe the excitations in Nd$_2$CuO$_4$ \cite{sachi}.

For the interactions between Cu spins ($H_{\mathrm Cu}$), we
employ the model used to describe the spin excitations of
antiferromagentic YBCO \cite{tranq}. This incorporates
nearest-neighbour couplings in-plane $J_\|$ and between planes
$J_\bot$, and an out-of-plane anisotropy. The optic mode gap at
the zone centre is $2\sqrt{J_\| J_\bot}$. For YBCO$_{6.2}$ the
value of $J_\|$, determined by Hayden {\itshape et al}
\cite{optics}, is $125\pm5$\,meV. Here, we use a value of
$J_\|=127\pm10$\,meV recently obtained on our crystal of
PrBCO$_{6.2}$ by inelastic neutron scattering \cite{maps},
consistent with Raman data \cite{yoshida}. Combined with our
measurement of the optic gap, this value of $J_\|$ leads to
$J_\bot=5.5\pm0.9$\,meV. We note the surprising result that the
value of $J_\bot$ for PrBCO$_{6.2}$ is a factor of 2 lower than
that for YBCO$_{6.2}$ ($J_\bot=11.5\pm1.5$ \cite{optics}).

Above $T_{\mathrm Pr}$ the Cu ordering is similar to that observed
in YBCO$_6$, however the presence of the interaction given by
(\ref{eq:Hcupr}) leads to a modification of the Cu magnetic
structure at temperatures below $T_{\mathrm Pr}$. The magnetic
moments on neighbouring Cu planes are observed to counter-rotate
about the $c$ axis by an angle $\phi$ \cite{ATBPRL97,incomm}. This
twist can be accounted for in our model by minimising the
contribution to the ground state energy from the $J_\bot$ and
pseudo-dipolar terms. The result is
\begin{equation}
\sin{\phi}=8\langle{J_x}\rangle\langle{S}\rangle K_{xy}/J_\bot
\label{eq:phi}
\end{equation}
where $J_x$ is the $x$-component of the Pr angular momentum. The
more general problem of verifying that the observed magnetic
structure minimises the energy of the combined Cu--Pr system has
been considered in \cite{maleev}.

Having modified $H_{\rm Cu}$ to take into account this ground
state configuration, the excitation spectrum of the complete
Hamiltonian $H=H_{\mathrm Cu}+H_{\mathrm Pr}+H_{\mathrm Cu-Pr}$ is
obtained from the bilinear terms in the spin-deviation operator
expansion in the usual way. As a result of the pseudo-dipolar
Cu--Pr interaction the spectrum is modified in two ways. First,
the bilayer twist $\phi$ leads to a gap in the in-plane acoustic
Cu mode (the out-of-plane mode is also raised and continues to lie
a few meV higher). Measurements below $T_{\mathrm Pr}$ at higher
energies indicate that the gap is $\simeq10$\,meV. Secondly, the
two Pr excitation branches are both significantly affected at the
magnetic zone centre by coupling with the Cu modes \cite{foot2}.
As would be expected, the Pr spectrum is little affected away from
the zone centres because of the steepness of Cu spin-wave branch.

Of the four independent components of the tensor {\bfseries K},
the major contribution at the zone centre is from $K_{xy}$. The
other components produce only relatively small effects, not
varying very strongly with $\bf Q$. Given that the Pr--Pr
interaction can account for the dispersion away from the zone
centre satisfactorily, we next fit the data with the dispersion of
the full model by varying the parameters $J_{\rm Pr}$ and
$K_{xy}$. A good agreement is obtained, as shown by the solid line
in Fig. 2 which corresponds to the values $J_{\rm Pr}=0.025$\,meV
and $K_{xy}=0.30$\,meV (the statistical errors in the fitted
parameters are better than 10\%), although near the nuclear zone
centre the predicted dispersion is so steep that the limited
resolution of the spectrometer will limit the extent to which the
model can be tested. Further support for the fitted value of
$K_{xy}$ comes from (\ref{eq:phi}) which, together with
$\langle{S}\rangle=1/2$ and $\langle{J_x}\rangle=1.9$ from the
single-ion model, gives $\phi=22^\circ$, in good agreement with
the observed value of $\phi=20\pm5^\circ$ \cite{ATBPRL97}.
However, we should also mention two remaining shortcomings of the
model. First, the fitted values for the exchange constants only
produce a molecular field at the Pr site about half the size of
that required to produce the observed energy splitting. Secondly,
the crystal field model predicts a Pr moment of $2\,\mu_{B}$
compared to the measured value of $1.2\,\mu_{B}$. These
discrepancies may indicate that the electronic state of the Pr ion
is not completely accounted for by the crystal field model, and
that other effects such as covalency need to be considered.

Before concluding, we consider the magnitude of $K_{xy}$ and
$J_{\rm Pr}$ in comparison with other related measurements.
Although no previous attempts have been made to determine the
Cu--R coupling in RBCO compounds, a similar pseudo-dipolar term
seems to be present in Nd$_2$CuO$_4$, for which a value of
$K_{xy}=0.075$\,meV was found, a factor 4 less than found here for
PrBCO \cite{sachi}. As for the size of $J_{\rm Pr}$, we can judge
this by making a simple estimate from mean-field theory of $T_{\rm
Pr}$ assuming $J_{\rm Pr}$ to be the only interaction present.
This requires a calculation of the single ion susceptibility using
the same crystal field parameters as previously and predicts
$T_{\rm Pr}\approx7$\,K. This value is surprisingly large,
suggesting that the high value of $T_{\rm Pr}$ observed
experimentally is due not only to a significant Cu--Pr interaction
but also to large value of the Pr--Pr exchange.

The central result of the work described in this paper is the set
of exchange constants $K_{xy}$, $J_{\rm Pr}$ and $J_\bot$,
determined for PrBCO for the first time. These three constants are
important because they reflect exchange pathways that depend on Pr
$4f$ states, and so contain information on the underlying
electronic structure of PrBCO. The $J_{\rm Pr}$, $K_{xy}$ are
found to be surprisingly large, and this shows that both the
Pr--Pr and Cu--Pr interactions are relatively strong in comparison
to the corresponding interactions in other cuprates. $J_\bot$, on
the other hand, is found to be a factor of two smaller in PrBCO
than in YBCO, indicating that Pr has a significant influence on
the interlayer exchange coupling. These results provide insight
into the anomalous electronic and magnetic properties of PrBCO,
and suggest that the calculation of exchange interactions in PrBCO
could be a very valuable test of models for its electronic
structure. Similar experiments on an oxidised crystal of PrBCO,
for which the effects of hybridisation could be even more
pronounced, are planned for the near future.

We acknowledge the financial support of the EPSRC, and the EU TMR
Programme enabling the experiments to be performed at Ris\o.



\begin{figure}
\begin{center}
\includegraphics[angle=0,clip=]{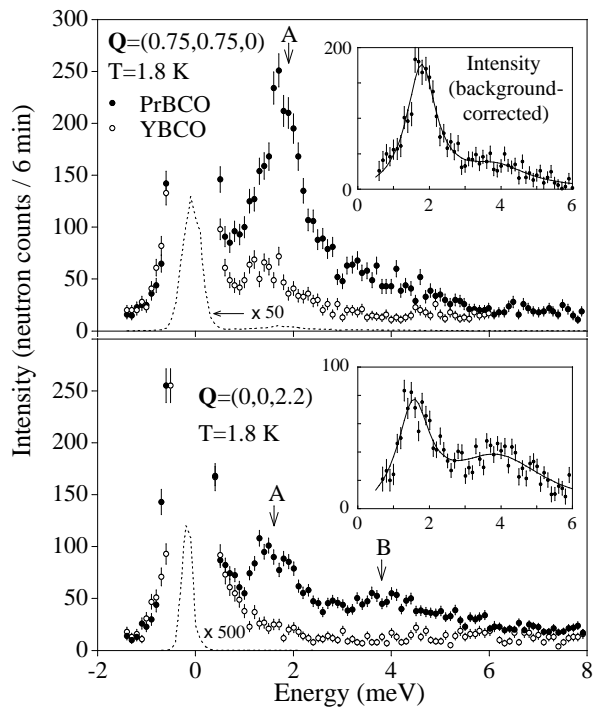} 
\caption{Neutron inelastic scattering from PrBCO and YBCO at 1.8
K, for {\bfseries Q}=$(0,0,2.2)$ and $(0.75,0.75,0)$. The elastic
peak from PrBCO is shown by the broken curve (reduced by a factor
of 50). Fits to the PrBCO data corrected with the YBCO as a
non-magnetic background is shown in the insets.}
\end{center}
\end{figure}
\begin{figure}
\begin{center}
\includegraphics[angle=0,clip=]{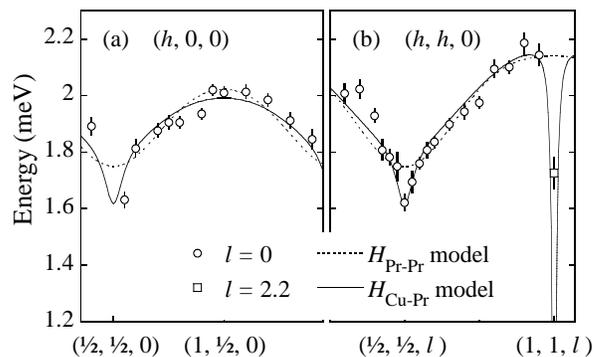} 
\caption{Dispersion of the low-energy Pr excitation. The measured
data, obtained by fitting as shown in Fig. 1, are plotted with
open symbols. The error bars are the standard deviations from the
fitting procedure, but in some cases the points are the result of
averaging several runs and the errors are then correspondingly
reduced. The broken line is a fit to a model for the Pr sublattice
only, while the solid line is a fit to the full model including a
Cu--Pr interaction. $(0.5,0.5,0)$ and $(1,1,0)$ correspond to the
centre of adjacent magnetic zones, so that the zone boundary is
crossed at $(0.75,0.75,0)$, but only the low energy mode has a
measurable intensity in the first zone, and the high-energy mode
in the second.}
\end{center}
\end{figure}
\begin{figure}
\begin{center}
\includegraphics[clip=]{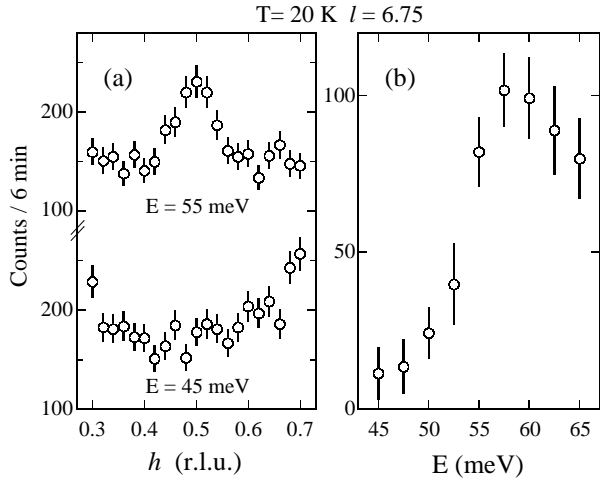}
\caption{(a) Measurement of the optic spin wave in PrBCO by
neutron inelastic scattering. Constant energy scans at 45 and
55\,meV through $(0.5,0.5,6.75)$ parallel to $(h,h,0)$ (in
reciprocal lattice units (r.l.u)) are shown. (b) The amplitude of
the observed peaks as a function of energy transfer.}
\end{center}
\end{figure}

\begin{references}
\bibitem{axes} The orthogonal axes ${x,y,z}$ are defined here with $x$
in the $ab$ plane of the crystal parallel to the in-plane
direction of the Pr moment, and $z$ along the $c$ axis.
\bibitem{foot2} From the structure of the eigenfunctions of the excitations, it
appears that the $K_{xy}$ component of the tensor couples the Cu
in-plane acoustic branch with the higher energy Pr excitation, and
the in-plane optic mode to the lower energy Pr excitation. Since
the out-of-plane modes do not couple to the Pr excitations, the
value of the out-of-plane anisotropy is not crucial here, so we
take the same value as found for YBCO.
\end{references}
\end{document}